\documentclass[a4paper]{jpconf}
\usepackage{graphicx}
\usepackage{amsfonts}
\usepackage{amsmath}
\usepackage{amssymb}

\begin{document}
\title{Constraining effective quantum gravity with LISA}

\author{Nicol\'as Yunes$^{1,2}$ and Lee Samuel Finn$^{1}$}
\address{$^{1}$ Department of Physics, The Pennsylvania State University, University Park, PA 16802, USA.}
\address{$^{2}$ Department of Physics, Princeton University, Princeton, NJ 08544, USA.}

\ead{LSFinn@psu.edu, nyunes@princeton.edu}

\begin{abstract}

All modern routes leading to a quantum theory of gravity -- i.e., perturbative quantum gravitational one-loop exact correction to the global chiral current in the standard model, string theory, and perhaps even loop quantum gravity -- require supplementing the Einstein-Hilbert action with a parity-violating Chern-Simons term. Such a term leads to amplitude-birefringent gravitational wave propagation: i.e., one (circular) polarization state amplified with propagation while the other is attenuated. The proposed Laser Interferometer Space Antenna (LISA)
is capable of observing gravitational wave sources at cosmological distances, suggesting the possibility that LISA observations may place a strong bound on this manifestation of quantum gravity. Here we report on a calculation of the effect that spacetime amplitude birefringence has on the signal LISA is capable of observing from inspiraling supermassive black hole binaries at large redshift. We find that the birefringence manifests itself in the observations as an anomalous precession of the binary's orbital angular momentum as it evolves toward 
coalescence, whose magnitude depends on the integrated history of the Chern-Simons coupling over the worldline of radiation wavefront. We estimate that LISA could place bounds on Chern-Simons modified gravity that are several orders of magnitude stronger than the present Solar System constraints, thus providing a probe of the quantum structure of spacetime. 
\end{abstract}

\section{Motivation}

Gravitational interactions described in General Relativity (GR) are symmetric under the discrete transformations of charge conjugation, parity- and time-reversal. All current routes to a quantum theory of gravity require the introduction of a parity-violating term to the Einstein-Hilbert action, which will only be apparent in the theory's gravitational sector. Experimental confirmation of gravitational symmetry breaking can serve as a guide to understand how the marriage of GR and quantum mechanics will take place. The observational details may even provide reason to choose among different approaches: e.g., different types of string theory or loop quantum gravity. Here we report on detailed calculations~\cite{Alexander:2007:gwp} of the effect this parity-violating correction has on gravitational wave propagation and how those effects manifest themselves in observations that the proposed Laser Interferometer Space Antenna (LISA)~\cite{Bender:1998,Danzmann:2003tv} will make possible. 

All modern approaches to a quantum gravitational theory of gravity lead to triangle-like anomalies, which must be canceled for mathematical consistency~\cite{AlvarezGaume:1983ig}. Cancellation of these anomalies leads to a Chern-Simons (CS) parity-violating contribution to the Einstein-Hilbert action. In four-dimensional compactifications of both perturbative and non-perturbative string theory, the inclusion of such a CS term is inescapable to cancel the Green-Schwarz anomaly and to satisfy duality symmetries in the presence of Ramond-Ramond scalars~\cite{polchinski:1998:stv,alexander:2006:css}. Parity-violation may also be present in loop quantum gravity if one requires invariance under large gauge transformations of the Ashtekar variables~\cite{Ashtekar:1989:tcp}, but the details of such a mechanism have not yet been worked out in detail. The effective quantum gravitational model that consists of the addition of such a CS term to the action goes by the name of CS modified gravity.

Gravitational experiments are the only route to test the CS extension, since this effective theory does not couple to photons. Solar system ephemerides can be used to place weak constraints on the magnitude of the CS coupling \cite{alexander:2007:npp,alexander:2007:ppe,smith:2007:eoc}, but gravitational wave observations have the potential to place much stronger bounds. This is because the CS parity-violation affects GW propagation by inducing an {\emph{amplitude birefringence}}~\cite{jackiw:2003:cmo,alexander:2006:css,alexander:2005:bgw}: i.e.~a polarization and wavelength-dependent amplification or attenuation of the GW amplitude, whose amplitude increases with distance propagated. The proposed Laser Interferometer Space Antenna \cite{Bender:1998,Danzmann:2003tv} --- a space-based gravitational wave detector --- is capable of observing inspiraling supermassive black holes as deep as redshift $30$ and over a period long enough that the radiation wavelength evolves over two orders of magnitude. The combination of propagation over cosmological distances and change of polarization amplitudes with signal wavelength offers the possibility of placing a strong bound on the magnitude of the CS coupling constant~\cite{Alexander:2007:gwp}. 

This paper is organized as follows: Section~\ref{CSBasics} describes the basics of CS modified gravity and its effect on GW propagation; Section~\ref{Observational} explains the observational implications of such an effect for GWs emitted by binary systems in the inspiral phase; Section~\ref{Fisher} presents a rough approximation of the bound that LISA could place on the CS coupling parameter given a GW detection; Section~\ref{conclusions} concludes and points to future research. In the remainder of this paper, we shall use the conventions of Misner, Thorne and Wheeler \cite{misner:1973:g}, described in more detail in~\cite{Alexander:2007:gwp}.

\section{Chern-Simons modified gravity}
\label{CSBasics}

The CS modification consists of enhancing the Einstein-Hilbert action through the addition of the product of a Pontryagin density term and a CS coupling function $\theta(x^{\alpha})$~\cite{jackiw:2003:cmo}. This effective theory yields the modified field equations 
\begin{equation}
\label{eq:EOM}
G_{\alpha\beta} + C_{\alpha\beta} = 8 \pi T_{\alpha\beta},
\end{equation}
where $G_{ab}$ is the Einstein tensor, $T_{ab}$ is the stress-energy tensor of the matter fields and $C_{ab}$ is a Cotton-like tensor, given by~\cite{jackiw:2003:cmo}
\begin{equation}
  C^{\alpha\beta} = 
  - \left[ 
  \theta_{,\epsilon}
    \epsilon^{\epsilon\gamma\delta(\alpha} \nabla_{\gamma} 
    R^{\beta)}{}_{\delta} - 
    \nabla_{\delta}\theta_{,\gamma} \;
    \epsilon^{\mu\nu\gamma(\alpha}{R^{\beta)\delta}}_{\mu\nu}  \right],
\end{equation}
$\theta$ is a function of spacetime coordinates, $\epsilon^{\alpha\beta\gamma\delta}$ is the Levi-Civita tensor density, and $R_{\mu \nu}$ is the Ricci tensor. Parenthesis in the superscript stand for symmetrization, commas in a subscript stand for partial differentiation and the $\nabla_{\mu}$ operator stands for covariant differentiation.

For gravitational wave perturbations propagating in a Friedmann-Robertson-Walker (FRW) background within CS modified gravity, the line element may be written as
\begin{equation}
\label{FRW}
ds^2 = a^2(\eta) \left[- d\eta^2 + \left(\delta_{ij} + h_{ij}\right) 
d\chi^i d\chi^j\right]
\end{equation}
where $\eta$ is conformal time, $\chi^i$ are comoving spatial coordinates, $\delta_{ij}$ is the Euclidean metric, $a(\eta)$ is the scale factor, and $h_{ij}$ is a transverse-traceless (TT) metric perturbation. Focusing attention on a plane-wave solutions in co-moving coordinates we introduce
\begin{equation}
\label{eq:ansatz}
h_{lm}(\eta,\chi^j) 
= \frac{\mathcal{A}_{lm}}{a(\eta)}
e^{-i\left[\phi(\eta) - \kappa n_k\chi^k\right]},
\end{equation}
where the amplitude $\mathcal{A}_{lm}$, the unit vector in the direction of wave propagation $n_k$ and the conformal wavenumber $\kappa>0$ are all constant. Resolving the amplitude into left- and right-handed polarizations simplifies the dispersion relation for the gravitational wave phase, which is found to satisfy
\begin{equation}
\label{eq:disp}
i\phi''_{\text{R,L}}+\left(\phi'_{\text{R,L}}\right)^2
+\mathcal{H}'+\mathcal{H}^2-\kappa^2=
\frac{i\lambda_{\text{R,L}}
\left(\theta'' - 2\mathcal{H} \theta'\right)
\left(\phi'-i\mathcal{H}\right)\kappa/a^2}%
{\left(1-\lambda_{\text{R,L}}\kappa \theta'/a^2\right)}
\end{equation}
where primes denote differentiation with respect to conformal time, ${\cal{H}} = a'/a$ is the conformal Hubble parameter and $\lambda_{\text{R,L}} = +1$ (or $-1$) for right- (or left-) polarized GWs.  

The evolution equation for the phase clearly indicates the characteristic signature of CS modified gravity. The left-hand side of Eq.~\eqref{eq:disp} is the familiar GR dispersion relation for wave propagation in FRW spacetimes, while the entirety of the CS correction appears on the right-hand side. This correction is a purely imaginary source term whose sign depends on the radiation wave polarization state and whose magnitude depends on the wavenumber and in particular on derivatives of the CS scalar. A critical aspect of this correction is that it will lead to a phase correction that depends on the entire integrated history of the CS coupling.

Treating the CS contribution to the dispersion relation as small, we can solve Eq.~\eqref{eq:disp} perturbatively. Assuming that $\theta'$ and $\theta'$  evolve on cosmological timescales, and that the present-day values of $\theta''_0/a_0^2$ and $\mathcal{H}_0\theta'_0/a_0^2$ are small, the accumulated CS correction to the plane wave phase in propagating from conformal time $\eta_i$ to the present is
\begin{eqnarray}
\Delta\phi_{1(\text{R,L})} &=& i \lambda_{\text{R,L}} \frac{k_0}{H_{0}} \xi(z),
\label{eq:dPhi1}
\end{eqnarray}
where we have defined the redshift-dependent CS form-factor $\xi(z)$ via 
\begin{equation}
\xi(z) = \frac{H_{0}^{2}}{2} \int_{0}^{z} dz \left(1 + z \right)^{5/2} \left[ \frac{7}{2} \frac{d\theta}{dz} + \left(1 + z\right) \frac{d^{2}\theta}{dz^{2}} \right],
\label{corr}
\end{equation}
and where $k_0 = a_{0} \kappa_{0}$ is very much greater than the Hubble constant ${H}_0 = \dot{a}/a$, with overhead dots standing for time differentiation and $z$ for redshift. 

The correction to the phase is purely imaginary with a sign that depends on the circular polarization state of the waves. Thus, one circular polarization state will be amplified and the other attenuated. The correction to the phase also depends linearly on the wavenumber -- it is stronger  for shorter wavelengths than for longer wavelengths. Finally, through Eq.~\eqref{corr}, its magnitude also depends on the integrated history of the CS coupling constant between the redshift of the source and the present. In the next section we will take advantage of all these effects to identify the signature of CS parity violation in the gravitational wave signal from inspiraling supermassive black hole binaries as observed by LISA. 

\section{Observational implications}
\label{Observational}

The proposed LISA gravitational wave detector will be sensitive to gravitational waves in the $(3\times10^{-5}, 100)$~mHz band, with peak sensitivity in the $(1,10)$~mHz band. Radiation from inspiraling black hole binaries with components of mass on order $10^6\,\mathrm{M}_\odot$ will radiate in this band during the final year before coalescence. The radiation itself will be nearly monochromatic, with its instantaneous radiation frequency increasing adiabatically over two orders of magnitude over this last year. As designed, LISA will be capable of detecting sources like these at redshifts up to approximately 30. 

Black hole binaries in the early inspiral phase will radiate gravitational waves which in CS gravity become
\begin{equation}
h_{\text{R,L}} = h_{\text{R,L}}^{GR}
\exp
\left[
 \lambda_{\text{R,L}} \frac{k_0}{H_{0}} \xi(z)
\right] \sim 
 h_{\text{R,L}}^{GR}
\left[1 + 
 \lambda_{\text{R,L}} \frac{k_0(t)}{H_{0}} \xi(z)
\right] 
\end{equation}
where $h_{\textrm{R,L}}^{GR}$ is the GR prediction for the right/left-polarized gravitational waveform. One can clearly see that the CS correction exponentially enhances or suppresses the gravitational wave amplitude depending on the polarization of the signal. Moreover, since the correction depends linearly on wavenumber, it also depends on time in a polynomial fashion.  

Focus attention on the ratio of the radiation amplitudes in the right and left circular polarization states: 
\begin{eqnarray}
\frac{h_{\text{R}}}{h_{\text{L}}}
&=& \frac{h_{\text{R}}^{GR}}{h_{\text{L}}^{GR}} \exp\left[\frac{2 k_{0}(t) \xi(z)}{H_{0}}\right] 
= \frac{1+x}{1-x}
\label{ratio}
\end{eqnarray}
where  $k_0(t)$ is the observed time-dependent radiation wavenumber (i.e., $f_0(t)/2\pi c$). In the absence of the CS correction the exponential term in $\xi$ is unity and the ratio defines the quantity $x$, which is, in fact, equal to $\cos\iota$, where $\iota$ is the direction angle between the line-of-sight to the binary system and the binary system's orbital angular momentum. 

In the presence of the CS correction the relationship between the observable $x$, defined by equation (10), and the actual inclination angle $\iota$ becomes 
\begin{equation}
x = \frac{\sinh\left(\frac{k_0(t) \xi(z)}{H_{0}} \right) + \cosh\left(\frac{k_0(t) \xi(z)}{H_{0}}\right) \hat{\chi}}{\cosh\left(\frac{k_0(t) \xi(z)}{H_{0}}\right) + \sinh\left(\frac{k_0(t) \xi(z)}{H_{0}}\right) \hat{\chi}} 
\sim \cos{\iota} + \frac{k_0(t) \xi(z)}{H_{0}}  \; \sin^{2}{\iota} + {\cal{O}}\left(\xi^{2}\right).
\end{equation}
If the binary components have significant spins then spin-orbit coupling will lead to a time-dependence of $\cos\iota$, and thus $x$. The effect of the CS coupling is to introduce \emph{apparent} evolution of $\cos\iota$, which tracks the radiation wavelength in a well-determined way. The time-dependence of this contribution is unique to the effects of the CS contribution, and thus, distinguishable from the physical precession of the binary system.  In the next subsection we estimate the magnitude of the CS correction that LISA may be able to observe. 

\section{Constraining quantum gravity with LISA}
\label{Fisher}

Given a GW detection with LISA from a supermassive black hole binary at large redshift, how well can we measure the CS correction? The answer to this question may be obtained by performing a full-co-variance matrix analysis that accounts for all eighteen parameters that characterize the binary system: eleven intrinsic ones (two component masses, two component spins and their orientation, the orbital phase, reference time when the phase, spins and eccentricity are measured), six extrinsic ones (two angles that define the orbital plane orientation, the source distance and angular location, the inclination angle), and one CS-related parameter $\xi$. This analysis would have to account for higher harmonics that arise due to higher-order contributions to the GW amplitude~\cite{arun:2007:hsh}, as well as non-vanishing eccentricity. Performing such a study is of course a formidable task that has not yet been carried out. 

We have made a series of plausible approximations to arrive at a crude estimate of the bound that LISA observations of an inspiraling binary black hole system could place on a CS correction to GR. First, we assume that we have two gravitational wave detectors, each of which observes a distinct polarization state of the radiation incident from the distant binary. Let the noise power spectral density in both detectors be white with magnitude $S_0$ across the entire, relevant band. Second, classify the eighteen parameters that characterize the binary into those that only determine the GW amplitude $\mu_{i}(t)$ (the distance, the inclination angle, and the CS parameter $\xi$), and those that affect only or principally the real part of the signal phase $\psi_{i}(t)$ (the orbital phase, the sky location, the binary period, and other parameters that describe the spins and angular momentum). Third, assume there is no correlation between these two sets of parameters, in which case the Fisher matrix becomes block-diagonal. In this case the inverse Fisher matrix is also block diagonal, with each block equal to the inverse of the corresponding block of the Fisher matrix. Finally, focus attention on the parameter block that includes the CS parameter $\xi$. 

For concreteness, let us now assume that there has been a GW detection from a non-precessing binary black hole system with known component masses $M = 10^{6} M_{\odot} (1 + z)^{-1}$, seen plane on ($\hat\chi_{0} = 0$) at a redshift of $z=15$. Such a system will produce a GW that, during the final year of observation, sweeps through the LISA band from a radiation wavelength of $c (10^{-4} \textrm{Hz} )^{-1}$ to $c (10^{-2} \textrm{Hz} )^{-1}$. Setting aside the antenna pattern functions, our approximate Fisher analysis suggests that the accuracy to which we could measure the CS parameter is given by
\begin{equation}
\delta \xi = \sqrt{\left(\boldsymbol{\Gamma}^{-1}\right)_{\xi \xi}} = 
1.7 \times10^{-20}\left(\frac{\sqrt{S_0}}{10^{-20}\,\text{Hz}^{-1/2}}\right)\left(1+z-\sqrt{1+z}\right),
\end{equation}
which corresponds to a ``$1$-sigma'' upper bound on $\xi$ of order $10^{-19}$. Such a result compares favorably, with Solar System experiments, which are orders of magnitude weaker and can constrain $\xi$ (and thus $\theta$) only locally.

Theory does not yet offer any real constraints on $\xi$: i.e., its magnitude depends strongly on the  theoretical framework that prescribes the function $\theta$. For example, in perturbative string theory, the theoretically expected magnitude of $\theta$ is on the order of the Planck scale. In the non-perturbative regime, however,  Ramond-Ramond charges will source the CS correction, leading to a significantly larger value~\cite{svrcek:2006:ais}. Even in perturbative frameworks, there are some models where $\xi$ could be enhanced, such as for late-time vanishing string couplings~\cite{brandenberger:1989:sie,%
tseytlin:1992:eos,%
nayeri:2006:pss,%
sun:2006:ccm,%
wesley:2005:cct,%
alexander:2000:bgi,%
brandenberger:2002:lpi,%
battefeld:2006:sgc,%
brandenberger:2006:sgc,%
brandenberger:2007:sgc,%
brax:2004:bwc}. 
An experimental measurement or bound on the magnitude of the CS coupling may thus select among alternative theoretical models for the quantum theory of gravity. 

\section{Conclusions}
\label{conclusions}

Parity violation is a defining characteristic of modern theoretical models for quantum gravity. One manifestation of this parity violation is a 
spacetime amplitude birefringence for gravitational (but not electromagnetic) wave propagation. LISA observations of the gravitational waves 
from cosmologically distant inspiraling black hole binary systems will show the effects of this spacetime birefringence as an apparent evolution of  the orbital angular momentum of the binary relative to the detector line-of-sight. This evolution has a unique time dependence that allows it to be distinguished from physical precession of the binary, which may also lead to an evolving inclination angle. We estimate that LISA observations are sufficiently sensitive to bound the parameter space of modern, theoretical models for quantum gravity and, thus, probe the fundamental nature of the universe. 

\ack 
This manuscript is based on~\cite{Alexander:2007:gwp}, which was written in collaboration with Stephon Alexander. We would also like to thank Ben Owen and Frans Pretorius for helpful discussions and comments. N.~Y.~acknowledges the support of the National Science Foundation (NSF) via grant PHY-0745779. LSF acknowledges the support of NSF awards PHY 06-53462 and PHY 05-55615, and NASA award NNG05GF71G. Lastly, we acknowledge the support of the Center for Gravitational Wave Physics, which is funded by the National Science Foundation under Cooperative Agreement PHY 01-14375.

\section*{References}
\bibliographystyle{iopart-num}
\bibliography{phyjabb,conf-proc}

\end{document}